\documentclass[a4paper, 10pt]{article}
\usepackage[utf8]{inputenc}
\usepackage{amssymb,amsmath,amsfonts,mathrsfs,pxfonts,latexsym,pgfplots,graphicx,graphics}
\usepackage{epstopdf,setspace}

\usepackage{pgfplots}
\pgfplotsset{compat=1.13}
\usepackage[mathlines]{lineno}
\usepackage[breaklinks=true,colorlinks=true,linkcolor=blue,urlcolor=blue,citecolor=blue]{hyperref}

\title{Combined variable field theory in FLRW cosmology}
\author{Avadhut Purohit}

\begin{document}

\maketitle
\begin{abstract}
  This work aims to study consequences of \textit{combined variable field theory} developed in [1] by analyzing FLRW models of gravity. It shows 
  $\Phi_{q \rightarrow 0} \rightarrow \infty$ is the Big bang in combined variable field theoretic settings. Nature of the Universe depends on the 
  new parameter $k$ appearing in the theory. Flat Universe has two disconnected branches namely expanding and collapsing Universe. The combined variable 
  field for closed Universe is complex in general. Whereas, the theory of open Universe is necessarily self-coupled combined variable field theory. But
  in the large $r$ limit, it reduces to the theory of flat Universe. 
  Resolution of the classical Big bang is a result of quantization program. Quantum theory predicts that the size of Universe was non-zero in the beginning.
   Combined variable field quantum in absence of scalar field coupling is interpreted as a quantum of space. The energy spectrum of this combined 
   variable field quantum varies with time.
\end{abstract}

\section{Introduction}
\hspace*{0.5 cm}
I will be working with massless scalar field throughout this paper. Theory formulated in [1] is recast into FLRW model of the Universe. So that 
consequences of the theory can be studied and results can be compared. The theory differs from other canonical theories such as loop quantum gravity and 
Wheeler-DeWitt theory. Unlike other canonical theories, $(\phi, q_{ab})$ are kinematical variables in combined variable field theory. The Universe is seen 
as a field. Similar to loop quantum cosmology, quantization of gravity is a result of quantization program and not an assumption. Theory suggests that 
the Universe began with finite, non-zero size. As Universe expanded, the energy of a quantum of space decreased sharply. In large $q$ limit, 
approaches to 3 units of Planck energy. \newline
\hspace*{0.5 cm}
When classical theory shows singularity, quantum theory expected to take over which would be free from those singularities. 
Therefore the Big bang singularity expected to get resolved in the quantum theory of gravity. There have been several attempts to quantize the gravity.
Loop quantum gravity which is also built under canonical formulation of general relativity is one of the promising theory among other theories. Loop 
quantum cosmology which is a branch of loop quantum gravity resolves the Big bang singularity. One can refer [2] in order to understand 
quantization program and resolution of the Big bang. An important lesson taken from loop quantum gravity is that the fields evolve with respect to one 
another. As mentioned in above paragraph, reason to choose massless scalar field because it acts as a clock. There are various papers suggesting this 
such as [2]. Whereas [3] discusses in detail about possibilities and limits under which scalar field can be used as a clock. These two points are key 
ingredients for this work. \newline
\hspace*{0.5 cm}
History taught us that quantizing non-relativistic theory and quantizing special relativistic theory is not equivalent. Quantization of non-relativistic 
theory happens to be straightforward whereas quantization of relativistic theory results in quantum field theory. When one classically evolves initial 
matter field configuration (which curves the space-time), underline space-time also evolves. This evolution is unique. But in case of quantum evolution, 
for a particular matter field configuration virtual multi-curvature states are possible. This very fact leads to combined variable field $\Phi$ 
distributed over metric and evolves with matter field. This is achieved by re-interpreting ($\hat{H} \Phi =0 $) $\Phi$ as a classical combined variable 
field (refer [1] for details). \newline
\hspace*{0.5 cm}
Section 2 is a brief overview of combined variable field theory formulated in [1]. Equations have been re-derived in the FLRW cosmological context. 
Hubble parameter can be obtained from Hamiltonian dynamics but it does not have natural combined variable field theoretic extension. Section 3 splits further
into three parts $\kappa=0$ (flat Universe), $\kappa =+1$ (closed Universe) and $\kappa=-1$ (open Universe). Section 4 contains quantum analog of section 3.
In the end, section 5 concludes earlier three sections. Throughout this paper I am going to work in units of $16\pi G =1$, $\hbar =1 $  and $c=1$. \newline
\section{Overview}
ADM formulation recasts gravity as a gauge field theory and thus brings closer to other special relativistic field theories. Zero Hamiltonian is a hallmark of 
background independent theories. 
\begin{equation} \label{eqn1}
 \int_{M} d^{3}x  \left( N^{a}H_{a} + |N| \left( H_{\text{scalar}} + H_{\phi} \right) \right) = 0  
\end{equation}
As mentioned in section 2, of paper [1], the full Hamiltonian constraints includes scalar Hamiltonian constraints, vector (or Diffeomorphism) 
constraints $H_{a}$ and matter field part. The scalar field taken here is a massless scalar field. The shift vector $N^{a}$ and lapse function $N$ are 
Lagrange multipliers. Shift vector together with the lapse function tells how neighbouring slices of hypersurfaces are deformed. 
\begin{equation} \label{eqn2}
 ds^2 = \left( N^2-N_{a}N^{a} \right) dt^2 -2N_{a} dt \hspace*{1 mm} dx^{a} - q_{ab}dx^{a}dx^{b}
\end{equation}
$q_{ab}$ is metric defined over 3 dimensional spatial manifold. By separating space and time dependent parts of gravitational field as well as scalar 
field ([1], section 2, (19)) and re-defining (rather a canonical transformation in the [1],section 2, (32)) , we get 
\begin{equation}\label{eqn3}
 H_{\text{total}}=\frac{1}{2}P^{2}_{\phi} - \frac{1}{2}\eta_{ij} P^{i}P^{j}+2 \alpha q_{k}P^{k} - V(\vec{q})
\end{equation}
scalar field potential is 0 in this case. $\eta_{ij}\coloneqq q_{i}q_{j}$. Combined variable field theory requires $\alpha=i$ and lapse function chosen
as (refer [1], section 2)
\begin{align} \label{eqn4}
 &  N^{a}=\frac{-\alpha}{\int_{M} d^3x \left( \mathring{q}_{ac}(\vec{x})D_{b}\mathring{P}^{bc}(\vec{x}) \right)}
 &N= \frac{\sqrt{\text{det } q_{a}(t)}}{\int_{M} d^3x  \left( \mathring{f}_{abcd}(\vec{x}) \mathring{P}^{ab}(\vec{x}) \mathring{P}^{cd}(\vec{x}) \right)}
\end{align}
Where $D_{b}$ is unique torsion-free covariant differential compatible with $q_{ab}$ and 
\begin{align} \label{eqn5}
 & \mathring{f}_{abcd} \coloneqq \frac{1}{\sqrt{\text{det } \mathring{q}_{ab}(\vec{x})}} \left( \mathring{q}_{ab}(\vec{x})
 \mathring{q}_{cd}(\vec{x})-
 \mathring{q}_{ac}(\vec{x})\mathring{q}_{bd}(\vec{x})-\mathring{q}_{bc}(\vec{x})\mathring{q}_{ad}(\vec{x}) \right) \\
 & \mathring{P}^{ab}(t,\vec{x}) \coloneqq \left(  \sqrt{\text{det }\mathring{q}_{ab}(\vec{x})}
 \left( \mathring{K}^{ab}(\vec{x}) - \mathring{K}(\vec{x}) \mathring{q}^{ab}(\vec{x}) \right) \right)
\end{align}
$\mathring{}$ indicates spatial component. $\mathring{K}_{ab}$ and $\mathring{K}$ are extrinsic curvature tensor and extrinsic curvature scalar respectively.
In absence of scalar field coupling combined field potential is purely gravitational in nature. Borrowing combined variable field potential from 
 (27), section 2, [1] 
\begin{equation} \label{eqn6}
   V(\vec{q}) =\text{det}(q_{a}(t)) \frac{\int_{M} d^{3}x \sqrt{\text{det } \mathring{q}_{ab}} \hspace*{0.1 cm} R^{(3)}(t,\vec{x})}{\int_{M} d^3x 
  \left( \mathring{f}_{abcd}(\vec{x}) \mathring{P}^{ab}(\vec{x}) \mathring{P}^{cd}(\vec{x}) \right)} 
\end{equation}
$R^{(3)}(t,\vec{x})$ is intrinsic curvature scalar in 3 dimensions. Spatial part of potential can be seen as a ratio of intrinsic spatial action to 
extrinsic spatial action. This form of combined potential is a result of separation of temporal and spatial dependence of field variables. \newline 
\begin{center}
  \textbullet FLRW model
\end{center}
3-Metric for FLRW models is given as
\begin{equation} \label{eqn7}
 q_{ab} \coloneqq q_{a}(t) \hspace*{1 mm} \text{diag}\left( \hspace*{1 mm} \frac{1}{1-\kappa r^2}, \hspace*{1 mm} r^{2}, 
 \hspace*{1 mm} r^{2} sin^{2} \theta \right)  
\end{equation}
$\kappa$ is 0 for flat Universe, +1 for closed Universe and -1 for open Universe.  Total Hamiltonian is obtained by using (39), section 2 of [1] for 
this FLRW metric 
\begin{equation} \label{eqn8}
 H_{\text{total}}=\frac{1}{2}P^{2}_{\phi} - \frac{1}{2}(3qP)^{2} + 6 \alpha qP - V(\vec{q})
\end{equation}
The third term is diffeomorphism term. In the paper [1], shift vector was chosen in such a way as to make combined variable field theory 
consistent. Here, only to make calculation easier I have chosen the right side of (25), section 2, [1] to be $\alpha$. Notice that the dynamics
depends neither on lapse function nor on the shift vector. Therefore, I choose $\alpha = 0$ to make calculation easier. But for combined variable 
field theory, the choice is given by (25), section 2, [1]. \newline
\hspace*{0.5 cm} Equations of motion for gravitational part are given by 
\begin{align} \label{eqn9}
 & \dot{q} \coloneqq \left\lbrace q, H_{\text{total}} \right\rbrace = - 9 q^{2}P  & \dot{P} \coloneqq
 \left\lbrace P, H_{\text{total}} \right\rbrace = - 9 qP^{2}
\end{align}
These are coupled equations in $q$ and $P$. First we obtained $P$ as a function of $q$ by taking ratio and integrating 
\begin{equation} \label{eqn10}
 P = \sqrt{q^{2}+ C}
\end{equation}
Finally, we get the solution to an equation of motion for $q$ by using above equation  
\begin{equation} \label{eqn11}
 q = \sqrt{C} \hspace*{0.1 cm} \text{tan}\left[\text{sin}^{-1}\left(\frac{1}{9C t}\right) \right]
\end{equation}
The domain of time is $t \in (\frac{1}{9C}, \infty)$. As time increases $q$ decreases and tends to zero in the limit $t \rightarrow \infty $.  
One can relate this result to standard cosmology using $d\tau = N(t) dt$. Invariant length element in standard cosmologies is 
$ds^{2} = d\tau^{2} - a(\tau)^{2}(f(r)dr^{2} + r^{2}d\Omega^{2}) $. Here, $\tau$ represents time used in the standard cosmologies. Also 
note that the scalar field of this theory is related to the scalar field used in the standard cosmologies by a canonical transformation given in [1], 
section 2, (32). This is the reason why the form of $q(t)$ and Hubble parameter ($\frac{\dot{a}}{a} = \frac{\dot{q}}{2q}$) is different from that of 
standard cosmologies. One can in principle work with the same scalar field (i.e. without canonically transforming it). In that case, combined variable 
theory developed could be different. Further more, whether that theory has same dynamics or not, remains a question. In fact, without that 
transformation, whether it has simpler combined variable field theoretic extension itself is a question. This goes beyond scope of this paper and explorations 
in that direction will be done in future work. \newline 
\hspace*{0.5 cm}Equations of motion for scalar field part are given by 
\begin{align}
  & \dot{\phi} \coloneqq \left\lbrace \phi, H_{\text{total}} \right\rbrace = P_{\phi} 
  & \dot{P_{\phi}} \coloneqq \left\lbrace P_{\phi} , H_{\text{total}} \right\rbrace = 0
\end{align}
$P_{\phi}$ is a constant of motion and therefore $\phi$ is a linear function of time. Notice that time `t' used in $q(t)$ is not a real 
physical time. Hamiltonian vanishes on the constraint surface. Therefore, Hamiltonian evolution is a gauge transformation and not a physical 
time evolution. Thus, evolution is better understood in terms of variation of $q$ with respect to the scalar field (i.e. $q(\phi)$). 
Above result shows that $q_{\phi \rightarrow \infty} \rightarrow 0$ and $q$ increases non-linearly with a decrease of scalar field $\phi$. 
In absence of a scalar field, variation of $q$ can not be realized (in isotropic space). \newline
\hspace*{0.5 cm} 
On re-interpreting ($\hat{H} \Phi = 0$) $\Phi$ as classical combined variable field (refer section 3, [1]), we get 
\begin{equation} \label{eqn12}
 \left(\partial^{2}_{\phi}-\partial^{i}\eta_{ij}\partial^{j}+V(q_{k},\phi)\right)\Phi(\phi,q_{k}) = 0
\end{equation}
Where $\eta_{ij} \coloneqq q_{i}q_{j}$ serves as a spatial metric for combined variable field theory. Note that the $H$ we used is the one with a correct 
 choice of shift vector used in (25), section 2, [1] and not $\alpha = 0$. For FLRW case, by using metric defined in (\ref{eqn7}) above combined 
 variable field equation is re-calculated as 
\begin{equation} \label{eqn13}
 \left( \frac{\partial^{2}}{\partial \phi^{2}} -9 q^{2}\frac{\partial^{2}}{\partial q^{2}} - 12 q \frac{\partial }{\partial q} + V(q)
 \right) \Phi (r, q, \phi) =0
\end{equation}
Stress-energy tensor (refer section 3, [2]) can be recalculated for FLRW theory. Energy density ($\rho$) and pressure ($p$) for combined variable field 
\begin{align} \label{eqn14}
 & \rho =T^{0}_{0} = \frac{1}{2} \left( (\partial_{\phi} \Phi)^{2} + \eta_{jk}\partial^{j}\Phi \partial^{k}\Phi + V(q) \Phi^{2} \right) \\ 
  p =-T^{1}_{1}= -T^{2}_{2} &=- T^{3}_{3} = \eta_{11}\partial^{1}\Phi \partial^{1}\Phi +
 \frac{1}{2} \left( (\partial_{\phi} \Phi)^{2} -\eta_{jk}\partial^{j}\Phi \partial^{k}\Phi - V(q) \Phi^{2} \right) \label{eqn42}
\end{align}
In the limit $q\frac{\partial \Phi}{\partial q} \rightarrow 0$,
\begin{align} \label{eqn15}
 & \rho =  \frac{1}{2} \left( (\partial_{\phi} \Phi)^{2} + V(q) \Phi^{2} \right) \\ \label{eqn16}
 & p  =  \frac{1}{2} \left( (\partial_{\phi} \Phi)^{2} - V(q) \Phi^{2} \right)
\end{align}
In the limit $\partial_{\phi} \Phi \rightarrow 0$ pressure becomes negative. \newline
\subsection{Flat Universe}
\hspace*{0.5 cm}
Solution to (\ref{eqn13}) is obtained using separation of variables. Let $\Phi \coloneqq T(\phi) Q(q) $,
\begin{align*}
 Q(q)\frac{\partial^{2}T(\phi)}{\partial \phi^{2}} = 9T(\phi) q^{2}\frac{\partial^{2}Q(q)}{\partial q^{2}} +
 12T(\phi) q \frac{\partial Q(q)}{\partial q}
\end{align*}
Dividing this equation by $T(\phi) Q(q)$ we get 
\begin{equation} \label{equation1}
 \frac{1}{T} \frac{\partial^{2} T}{\partial \phi^{2}} = k^{2}=\frac{9 q^{2}}{Q} \frac{\partial^{2} Q}{\partial q^{2}} +
 \frac{12 q}{Q} \frac{\partial Q}{\partial q} 
\end{equation}
Left side of equation is purely scalar field dependent and right side is purely gravitational field dependent. Parameter $k$ therefore must be a constant. 
Solution to the gravitational part as well as scalar field part is given respectively as
\begin{align*}
 & Q(q) = q^{-\frac{1}{6}} \left( C_{1} q^{\frac{1}{6}\sqrt{1+4k^{2}}} + C_{2} q^{-\frac{1}{6}\sqrt{1+4k^{2}}} \right)  \text{\hspace*{0.5 cm} and}
& T(\phi) =  e^{k \phi} + B_{2}\hspace*{1 mm} e^{-k \phi} 
\end{align*}
\begin{equation}
 \Phi (\phi, q) = q^{-\frac{1}{6}} \left( C_{1} q^{\frac{1}{6}\sqrt{1+4k^{2}}} + C_{2} q^{-\frac{1}{6}\sqrt{1+4k^{2}}}
 \right) \left(  B_{1} \hspace*{1 mm} e^{k \phi} + B_{2}\hspace*{1 mm} e^{-k \phi} \right)
\end{equation}
Constants $k, C_{1},C_{2},B_{1}$ and $B_{2}$ can be settled by applying appropriate boundary conditions.  
Combined variable field has singularity at $q=0$ indicating the Big bang. \newline
For $\sqrt{1+4k^{2}}>1$: Combined variable field has two branches. The first, indicating collapsing Universe and
the second indicating expanding Universe. \newline
For $\sqrt{1+4k^{2}}= 1$: The first branch is steady state Universe whereas second is expanding Universe. \newline
For $0\leq \sqrt{1+4k^{2}}< 1$: There exists only expanding Universe. \newline
For $\sqrt{1+4k^{2}} \in \mathbb{I}^{+} $: Combined variable field is complex. This can be seen as linear combination of two real scalar
combined variable fields. \newline 
\hspace*{0.5 cm}
The left side of (\ref{equation1}) is 
\begin{equation}
  \frac{1}{T} \frac{\partial^{2} T}{\partial \phi^{2}} = k^{2}
\end{equation}
It is already shown in section 2 that scalar field is a linear function of time. Therefore $T(\phi)$ of combined variable field is equivalent to
 $\phi(t)$ of a scalar field. \newline \newline
\textbullet \textbf{Remarks:} \newline 
\hspace*{0.5 cm}
For $k^{2}< -\frac{1}{4}$, complex nature of combined variable field possibly indicating the existence of two independent real combined variable fields or 
in other words existence of two independent Universes. For $k^{2} > -\frac{1}{4}$, there exists just one combined variable field. If one demands the existence of
just our Universe then this condition puts a restriction on parameter $k$. 
$\lim_{q\rightarrow 0} \Phi = \infty$ is the Big Bang singularity. This parameter $k$ decides nature of the Universe. 
Expansion or collapse of the Universe is a result of $\Phi \rightarrow 0$. Using (\ref{eqn15}) and (\ref{eqn16}) we can also see that the combined 
variable field satisfies $\rho = p$ in the limit $q\frac{\partial \Phi}{\partial q} \rightarrow 0$.

\subsection{Closed Universe}
\hspace*{0.5 cm}
For closed Universe (i.e. $\kappa = +1$), spatial part of the metric is $q_{ab} = \text{diag}\left( \frac{1}{1 - r^2}, r^2, r^2 \text{sin}^{2}\theta 
\right) $. Therefore, intrinsic curvature is positive i.e. $R^{(3)}=\frac{6}{q}$. Using unit vector $n_{a}$ normal to the surface  $r=\textit{constant}$
spatial part of extrinsic curvature tensor and extrinsic curvature scalar can be calculated. 
\begin{align*}
  K_{ab} \coloneqq D_{a}n_{b} =  &\text{diag} \left( -r \sqrt{\frac{1}{1-r^2}}, \frac{r-r^3}{\sqrt{\frac{1}{1-r^2}}}, 
  -\frac{r \left(-1+r^2\right) \text{sin}^{2}\theta }{\sqrt{\frac{1}{1-r^2}}}  \right) \\
  & K=\frac{2-3 r^2}{r \sqrt{\frac{1}{1-r^2}}}
\end{align*}
These curvatures are defined on 3-manifold which can be pulled back to 4-dimensional spacetime manifold using projections, lapse function and shift vector.
Spatial part of combined potential is a ratio of intrinsic action to extrinsic action as shown in (\ref{eqn6}). 
\begin{equation}
  V(r,q)= 2\left( \frac{2880 \left(-r+r^3+\sqrt{1-r^2} \text{sin}^{-1}r\right)}{\sqrt{1-r^2} \left(r \sqrt{1-r^2} 
 \left(3105-6890 r^2+8312 r^4-4944 r^6+1152 r^8\right)+735 \text{sin}^{-1}r \right)} \right) q^{2}
\end{equation}
\begin{center}
 \includegraphics{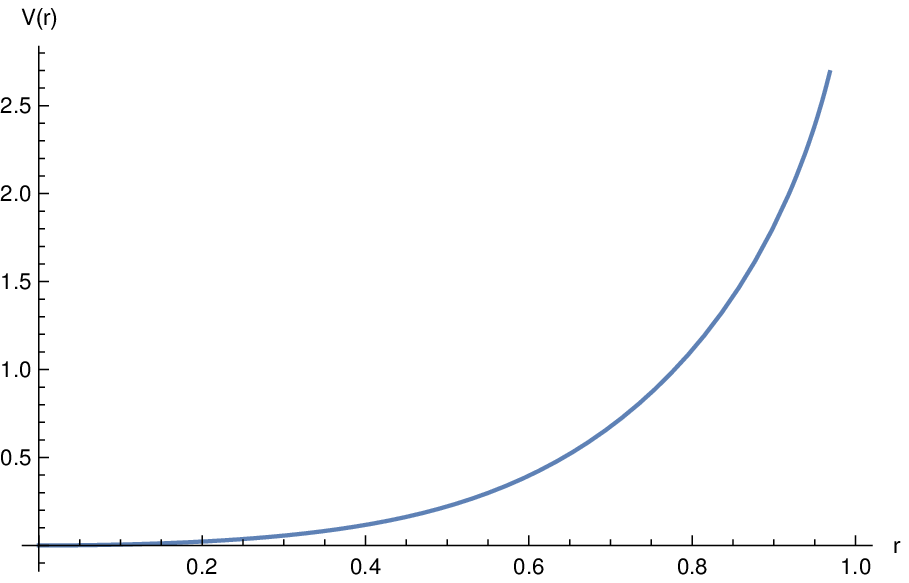}
\end{center}
\hspace*{0.5 cm}
Combined potential varies quadratically with $q$ having positive $r$ dependent coupling. This coupling increases with $r$. On implementing energy 
conservation (which $\int d^{3}q \hspace*{0.1 cm} \rho$ where $\rho$ is energy density of the combined variable field) one can infer that the combined 
variable field behaves as an oscillator. Solution to the equation of motion 
\begin{equation}
 \frac{1}{T} \frac{\partial^{2} T}{\partial \phi^{2}} = k^{2}=\frac{9 q^{2}}{Q} \frac{\partial^{2} Q}{\partial q^{2}} +
 \frac{12 q}{Q} \frac{\partial Q}{\partial q} - 2 V(r) q^{2}
\end{equation}
is given as 
\begin{align}
 \Phi (\phi, q) = & C_{1} J\left( \frac{\sqrt{1+4 k^{2}}}{6}, -i\frac{q \sqrt{2V}}{3} \right)  \left(  B_{1} \hspace*{1 mm} e^{k \phi} 
 + B_{2}\hspace*{1 mm} e^{-k \phi} \right) \\ \nonumber + 
  & C_{2} Y\left( \frac{\sqrt{1+4 k^{2}}}{6}, -i\frac{q \sqrt{2V}}{3} \right)  \left(  B_{1} \hspace*{1 mm} e^{k \phi} 
 + B_{2}\hspace*{1 mm} e^{-k \phi} \right)
\end{align}
\textbullet \textbf{Remarks:} \newline 
\hspace*{0.5 cm}
The combined variable field for closed Universe is complex in general. This can also be seen as a linear combination of two independent real
combined variable fields. By analyzing in further detail, one can find out the nature of closed Universe depending on the initial conditions and
the parameter $k$. This will be done in the future work. Equations (\ref{eqn15}) and (\ref{eqn16}) suggest that near $r=0$, $\rho=p$ and near 
$r=1$,  $\rho \approx - p$.
\subsection{Open Universe}
\hspace*{0.5 cm}
For open Universe (i.e. $\kappa = -1$), spatial part of the metric is $q_{ab} = \text{diag}\left( \frac{1}{1 + r^2}, r^2, r^2 \text{sin}^{2}\theta 
\right) $. Therefore, an intrinsic curvature is positive i.e. $R^{(3)}=-\frac{6}{q}$. The spatial part of extrinsic curvature tensor 
as well as extrinsic curvature scalar are given as
\begin{align*}
 K_{ab}= \text{diag} & \left( r \sqrt{\frac{1}{1+r^2}}, \frac{r}{\left(\frac{1}{1+r^2}\right)^{3/2}},
 \frac{r \hspace*{1 mm} \text{sin}^{2}\theta}{\left(\frac{1}{1+r^2}\right)^{3/2}} \right) \\
 & K= \frac{2+3 r^2}{r \sqrt{\frac{1}{1+r^2}}}
\end{align*}
Then combined variable field potential is obtained using these curvatures. 
\begin{equation}
 V(r,q)=-2\left( \frac{2880 \left(r+r^3-\sqrt{1+r^2} \text{sinh}^{-1}r\right)}{3105 r+9995 r^3+15202 r^5+13256 r^7+6096 r^9+1152 r^{11}+735 \sqrt{1+r^2}
 \text{sinh}^{-1}r} \right) q^{2}
\end{equation}
\includegraphics{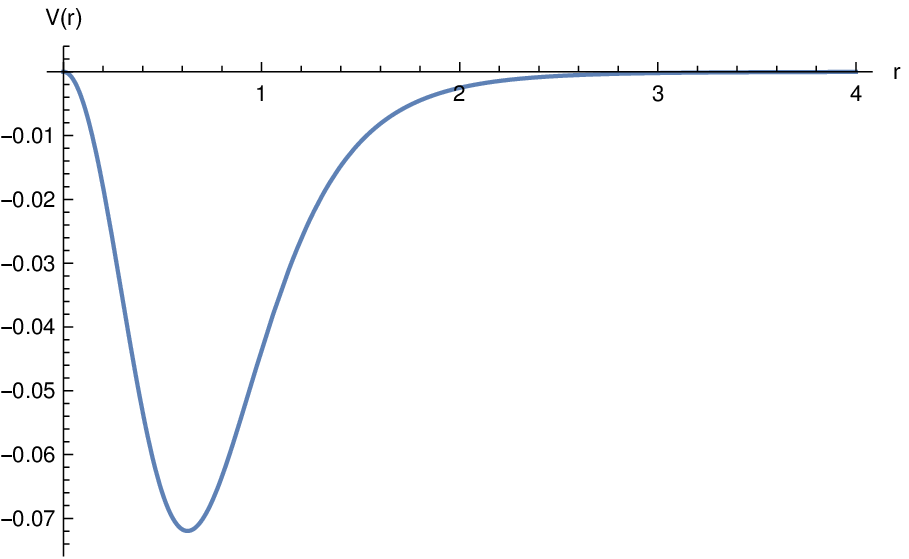} \newline
\textbullet \textbf{Remarks:} \newline \newline
This combined potential is also quadratic in $q$ but comes with a negative signature. As mentioned in [1] for negative potential $\Phi = 0$ is not a minima. 
Instead $\Phi_{vac} = \pm \sqrt{\frac{-V(r,q)}{\alpha}}$ is a minima. where $\alpha$ is self-coupling constant. Therefore for negative spatial 
curvature, theory is a self-coupled combined variable field theory. But in the large $r$ limit, theory becomes a free combined variable field theory or 
the theory of flat Universe. 


\section{Quantum Theory}
\subsection{Overview}
Quantization program of [1], section 3.2 shows that the Hamiltonian operator for free combined variable field can be written in terms of 
collection of an infinite harmonic oscillators and $| \phi, \vec{q} \rangle $ are eigen states of the Hamiltonian. 
\begin{equation} 
 \hat{\textbf{H}} =  \int d^{D}q \hspace*{1 mm}\left| \omega(\phi,\vec{q}) -D \right| \hat{n} 
  + \int d^{D}q  \frac{1}{2} \hspace*{0.1 cm} \omega(\phi,\vec{q}) 
 \hspace*{0.1 cm} \delta(\vec{0}) 
\end{equation}
$D$ is dimensional parameter depends on how many components of the metric are time dependent. $\omega (\phi, \vec{q})$ is a solution to Riccati equation 
appearing at the end of this subsection. Unlike standard Hamiltonians, this Hamiltonian gives $\phi$ evolution. At first glance, it may sound weird but
gravity being a dynamical theory of space-time it does not evolve with respect to any external time. Instead, it evolves with respect to a scalar field.
This Hamiltonian is a function over the phase space ($\Phi$, $\Pi$). The only non-trivial commutation relation (refer (57) section 3.2, [1]) between 
these fields is given by 
\begin{equation}
 \left[ \hat{\Phi} (\phi,\vec{q}) , \hat{\Pi} (\phi, \vec{q}^{\prime}) \right] = i \delta (\vec{q},\vec{q}^{\prime})
\end{equation}
and only non-trivial commutation relation (refer (64) section 3.2, [1]) between creation and annihilation operators satisfy 
 \begin{equation}
  \left[ \hat{a} (\phi,\vec{q}),\hat{a}^{\dagger}(\phi,\vec{q}^{\prime}) \right] = \left(\omega (\phi ,q) - D\right) \delta (\vec{q},\vec{q}^{\prime})
 \end{equation}
 $D$ is a number of spatial dimensions. A number operator is defined as $\hat{n} \coloneqq \hat{a}^{\dagger}\hat{a} $ for $\omega - D >0$
 whereas for $\omega - D <0$ it is $\hat{n} \coloneqq \hat{a} \hat{a}^{\dagger} $. This is because of role of creation and annihilation operator gets 
 reversed in these two domains. $\omega$ is a solution to following Riccati equation (refer (62), section 3.2, [1])
 \begin{equation} 
    V(\phi, \vec{q}) = \omega^{2}-\vec{\frac{\partial}{\partial q}}.\left( \vec{q}\omega \right)  = \omega^{2} - D \omega - 
   \vec{q}.\frac{\partial \omega}{\partial \vec{q}}
 \end{equation} 
 This equation arises in the process of quantization. Refer to quantum theory section 3.2 of [1] for more details.  

\subsection{Flat Universe}
\hspace*{0.5 cm}
For flat (3 dimensional) Universe with a massless scalar field, the solution to Riccati equation
\begin{equation} 
 \omega^{2} - 3 \omega - 3q\frac{\partial \omega}{\partial q} =0 
\end{equation}
\begin{equation}
 \omega = \frac{3}{1+ A q}
\end{equation}
Where $A$ is constant which can be fixed by applying proper boundary conditions. Hamiltonian operator for flat Universe with massless scalar field is 
\begin{equation}
 \hat{\textbf{H}} =  \int d^{3}q \hspace*{1 mm}\left| \frac{3}{1+ A q} -3 \right| \hat{n} 
  + \int d^{3}q  \frac{1}{2} \hspace*{0.1 cm} \frac{3}{1+ A q} \hspace*{0.1 cm} \delta(\vec{0}) 
\end{equation}
\begin{center}
 \includegraphics{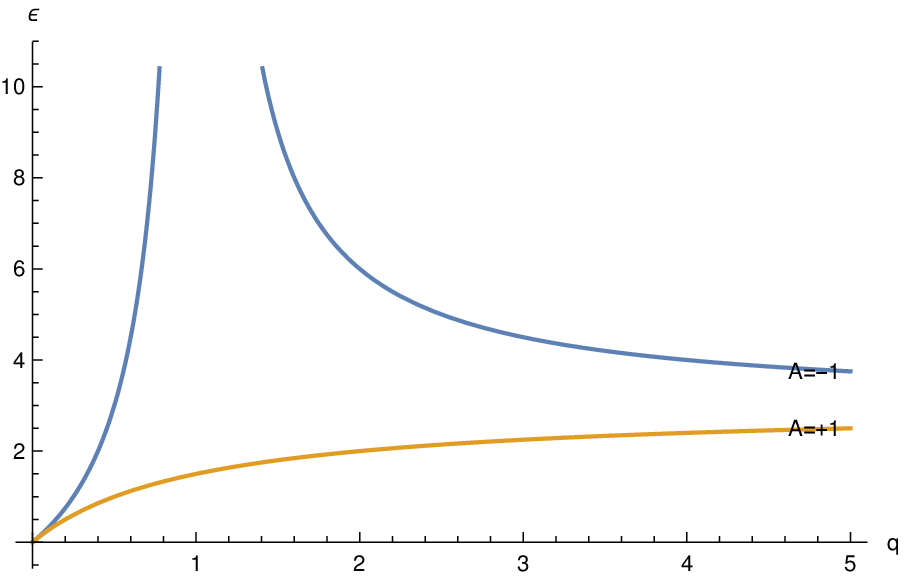}
\end{center}
For $A=0$: $\epsilon =0$ the theory is classical and does not have quantum analog.  \newline
For $A>0$: $\epsilon$ increases with $q$ and approach to 3. Classical nature of the Universe at the big bang (at $q = 0$)
 which later becomes more and more quantum makes this condition unphysical. \newline
For $A<0$: $ \lim_{q \rightarrow \infty} \epsilon \rightarrow 3$. Diverging nature of $\epsilon$ for $q\neq 0$ tells us 
that the Universe did not begin at $q=0$ instead the Universe began at $q\neq 0$ ( $=q_{0}$). Constant $A$ therefore can be identified as 
$A=-\frac{1}{q_{0}}$. \newline 
\hspace*{0.5 cm} Only $A<0$ is a physical choice. Because for $A=0$, the Universe is completely classical and in case of $A>0$, the 
Universe is classical at $q=0$ (i.e. at the Big Bang). 
$A=-1/q_{0}$ has two branches $q<q_{0}$ and $q>q_{0}$. If in the beginning our Universe had $q < q_{0}$, the Universe would have been 
collapsing Universe. But for $q>q_{0}$, the Universe is expanding. As $q$ increases, the energy of the combined variable field quantum 
decreases and approaches to 3 units of Planck energy. The Hamiltonian operator is a collection of an infinite combined variable field
 quantum. Creation operator acting on the vacuum state $|0 \rangle $ produces $|1,q \rangle$. Quantum vacuum (which is infinite!) may 
thought to be a sea of constantly creating and annihilating combined variable field quantum. \newline

\subsection{Closed Universe}
\hspace*{0.5 cm}
$\epsilon (r,q)$ is obtained using solution to Riccati equation (refer (62), section 3.2, [1]) as shown below
\begin{equation} \label{eqn44}
 \omega^{2} - 3 \omega - 3q\frac{\partial \omega}{\partial q} = V(r) q^{2} 
\end{equation}
\begin{equation} \label{eqn45}
 \omega (\phi,q) = \frac{\left(3 A+ q \sqrt{-3V}\right) \sin \left(q \sqrt{-\frac{V}{3}}\right)+\left(3-A q \sqrt{-3V}\right) 
 \cos \left(q \sqrt{-\frac{V}{3}}\right) }{A \sin \left(q \sqrt{-\frac{V}{3}}\right)+\cos \left(q \sqrt{-\frac{V}{3}}\right)}
\end{equation}
\begin{center}
\includegraphics{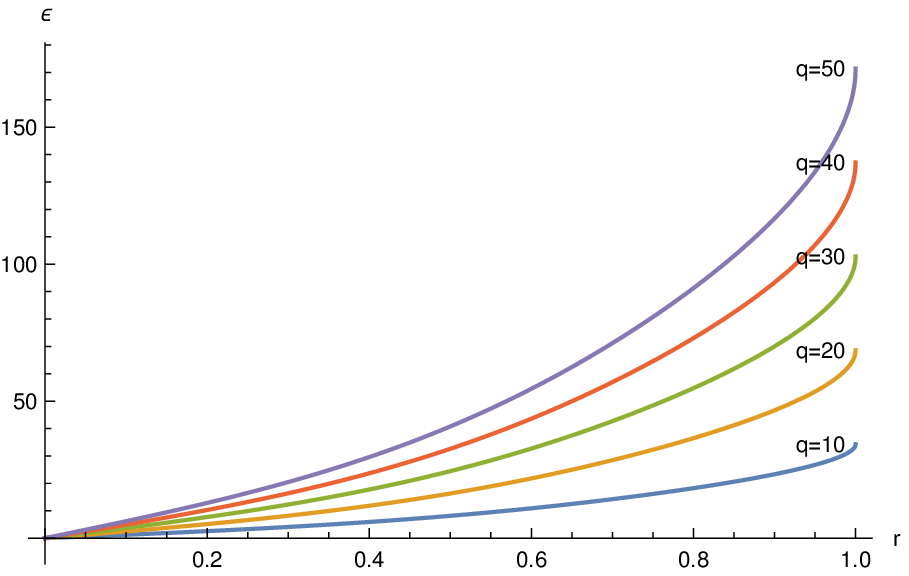}
\end{center}
Above plot of $\epsilon (r,q)$ verses $r$ shows that the combined variable field quantum have different energy spectrum at different 
$r$ and at different times as well (through time-dependent fields). $\epsilon$ hits singularity at $r=0$. Above result is true for any value of 
$A$. 


\subsection{Open Universe}
\hspace*{0.5 cm}
As mentioned in the classing part of the theory, this theory is self-coupled combined variable field theory. Combined potential in this case is 
$\frac{1}{2} V(r,q) \Phi^{2} + \frac{1}{4} \alpha \Phi^{4}$. But I have taken a zeroth order approximation of the theory in order to understand zeroth 
order quantum effects. This approximation is sensible because combined potential is negligible in large $r$ limit and therefore theory (at least
classically) becomes equivalent to that of the theory with flat Universe.

\begin{center}
 \includegraphics{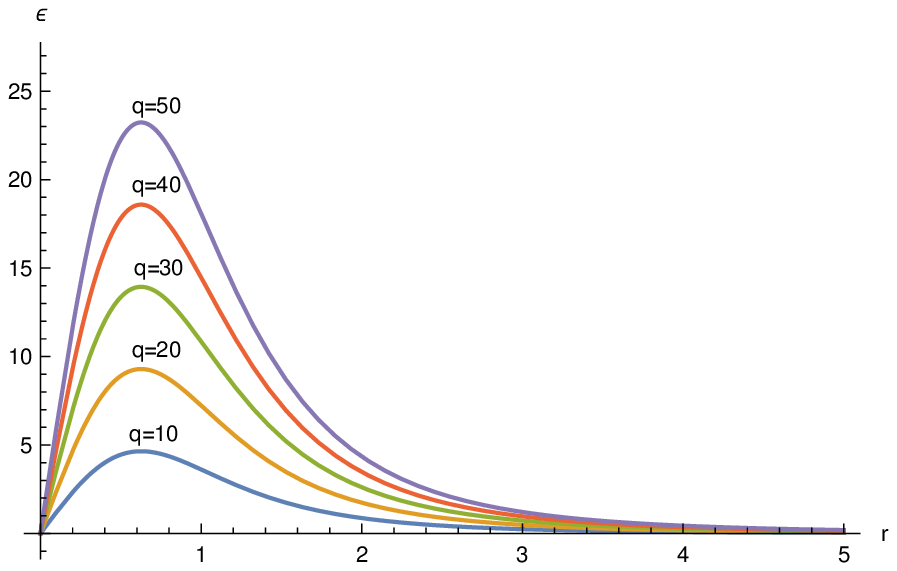}
\end{center}
The form of  $\omega$ required is same as that of given in (\ref{eqn45}) but combined potential is different. 
In the limit $r \rightarrow \infty$, the theory becomes a classical theory (as $\epsilon_{r \rightarrow \infty} \rightarrow 0$).
The role of creation and annihilation operator is exactly opposite compared to that of the theory with $\kappa = +1$. 
Detail analysis of this theory will be carried out in the next work.

\section{Conclusion}
\hspace*{0.5 cm}
Classical theory shows nature of the Universe can be expanding, static or oscillatory depends upon the parameter $k$. Further investigation is required
in order understand the physics of parameter $k$. $\lim_{q=0} \Phi \rightarrow \infty $ is the combined variable field theoretic understanding of the 
Big Bang. This theory does not have direct analog of Hubble parameter because metric field and scalar field are not observables. 
Expansion of the Universe is a result of the combined variable field $\Phi \rightarrow 0$ or energy density $\rho$ of combined
variable field approaching 0. In case of flat Universe, similar to other standard cosmologies this theory also has two disconnected branches, expanding 
and collapsing branch. THe combined variable field for closed Universe is complex and can be thought of a linear combination of two independent real combined
variable fields. Whereas, the theory of open Universe is self-coupled combined variable field theory. Further detail analysis of these two theories will 
be done in the future work.
The equation of state for flat Universe in the large $q$ limit shows combined variable field behaves as a normal scalar field. The equation 
of state for closed Universe shows both solutions normal and dark energy. Near $r=0$ it behaves as a normal field but near $r=1$ it behaves as a dark
energy field. The equation of state for open Universe, in general, behaves as a normal field. \newline
\hspace*{0.5 cm}
Quantum theory of flat Universe resolves the Big Bang singularity through quantum dynamics. It suggests that the Universe began with finite and non-zero 
size with $q_{0}$ being metric at the beginning. The spectrum of combined variable field quantum which is interpreted as a quantum of space had very 
high energy. It then decreased and in the limit $q \rightarrow \infty$ it approaches to $3E_{\text{Planck}}$. This interpretation of combined variable 
field quantum being a quantum of space is not possible in presence of scalar field couplings. In general quantum of combined variable field is 
 neither a quantum of space nor a quantum of scalar field. It is a quantum of both combined together. Quantum theory of closed and open Universe shows 
that the energy of a quantum of space varies also with $r$. \newline

\section{References}
$[1]$ \hspace*{0.2 cm} Canonical gravity and scalar fields - Avadhut Purohit \hspace{0.2 cm} arXiv:1704.04004v3 [gr-qc] 02 Feb 2018 \newline \newline
$[2]$ \hspace*{0.2 cm} Quantum Nature of the Big Bang - Abhay Ashtekar, Tomasz Pawlowski, and Parampreet Singh \hspace*{0.2 cm} arXiv:gr-qc/0602086v2 6
 Apr 2006 \newline \newline
$[3]$ \hspace*{0.2 cm} Scalar field as a time variable during gravitational evolution - Anna Nakonieczna and Jerzy Lewandowski \hspace*{0.2 cm} 
 arXiv:gr-qc/1508.05578v2 24 Sep 2015 \newline \newline 

\section{Acknowledgment}
I am immensely grateful to Prof. Ajay Patwardhan for providing expertise. This work would not have possible 
without his guidance. Any errors are my own and should not tarnish his reputation.

\end{document}